\documentclass[12pt]{article}
\usepackage{amsfonts}
\newcommand{\rpar}{\stackrel{\leftarrow}{\partial}}
\newcommand{\lpar}{\stackrel{\rightarrow}{\partial}}
\begin{document}
\small
\renewcommand{\thefootnote}{\fnsymbol{footnote}}
\begin{center}
{\large\bf Superspace Generalizations
of Schouten-Nijenhuis Bracket\\}
\vspace{0.5cm}
Dmitrij V. Soroka\footnote{E-mail: dsoroka@kipt.kharkov.ua} and
Vyacheslav A. Soroka\footnote{E-mail: vsoroka@kipt.kharkov.ua}
\vspace{0.5cm}\\
{\it Kharkov Institute of Physics and Technology,
61108 Kharkov, Ukraine}\\
\vspace{0.5cm}
\end{center}
\begin{abstract}
{The Schouten-Nijenhuis bracket is generalized for the superspace case
and for the Poisson brackets of opposite Grassmann parities.}
\end{abstract}
\renewcommand{\thefootnote}{\arabic{footnote}}
\setcounter{footnote}0
\bigskip


{\bf I.\/} Recently a prescription for the construction of new Poisson brackets from the 
bracket with a definite Grassmann parity was proposed 
\cite{Soroka:soroka&soroka}. This 
prescription is based on the use of exterior differentials of diverse 
Grassmann parities. It was indicated in~\cite{Soroka:soroka&soroka} that this 
prescription leads 
to the generalizations of the Schouten-Nijenhuis bracket~
\cite{scho,nij,nij1,fr-nij,kod-sp,but,bffls,oz,karmas} on the both superspace case 
and the case of the brackets with diverse Grassmann parities. In the present 
report we give the details of these generalizations\footnote{Concerning the 
generalizations of the Schouten-Nijenhuis bracket see also~\cite{az1,az2}.}.


{\bf II.\/} Let us recall the prescription for the construction from a given Poisson 
bracket of a Grassmann parity $\epsilon\equiv0,1\pmod2$ of another one.

A Poisson bracket, having a Grassmann parity $\epsilon$, written in arbitrary
non-canonical phase variables $z^a$
\begin{eqnarray}\label{2.1}
\{A,B\}_\epsilon=A\rpar_{z^a}\omega_\epsilon^{ab}(z)\lpar_{z^b}B,
\end{eqnarray}
where $\rpar$ and $\lpar$ are right and left derivatives respectively,
has the following main properties:
\begin{eqnarray}
g(\{A,B\}_\epsilon)\equiv g_A+g_B+\epsilon\pmod2,\nonumber
\end{eqnarray}
\begin{eqnarray}
\{A,B\}_\epsilon=-(-1)^{(g_A+\epsilon)(g_B+\epsilon)} \{B,A\}_\epsilon,
\nonumber
\end{eqnarray}
\begin{eqnarray}
\sum_{(ABC)}(-1)^{(g_A+\epsilon)(g_C+\epsilon)} 
\{A,\{B,C\}_\epsilon\}_\epsilon=0,\nonumber
\end{eqnarray}
which lead to the corresponding relations for the matrix
$\omega_\epsilon^{ab}$
\begin{equation}
g\left(\omega_\epsilon^{ab}\right)\equiv g_a+g_b+\epsilon\pmod2,\label{2.2}
\end{equation}
\begin{equation}
\omega_\epsilon^{ab}=-(-1)^{(g_a+\epsilon)(g_b+\epsilon)} 
\omega_\epsilon^{ba},\label{2.3}
\end{equation}
\begin{equation}
\sum_{(abc)}(-1)^{(g_a+\epsilon)(g_c+\epsilon)} 
\omega_\epsilon^{ad}\partial_{z^d}\omega_\epsilon^{bc}=0,\label{2.4}
\end{equation}
where $\partial_{z^a}\equiv\partial/\partial z^a$ and $g_a\equiv g(z^a)$, 
$g_A\equiv g(A)$ are the corresponding Grassmann parities of phase coordinates
$z^a$ and a quantity $A$ and a sum with a symbol $(abc)$ under it
designates a summation over cyclic permutations of $a, b$ and $c$.

The Hamilton equations for the phase variables $z^a$, which correspond
to a Hamiltonian $H_\epsilon$ ($g(H_\epsilon)=\epsilon$),
\begin{eqnarray}
\frac{dz^a}{dt}=\{z^a,H_\epsilon\}_\epsilon=
\omega_\epsilon^{ab}\partial_{z^b}H_\epsilon\label{2.5}
\end{eqnarray}
can be represented in the form
\begin{eqnarray}
\frac{dz^a}{dt}=\omega_\epsilon^{ab}\partial_{z^b}H_\epsilon\equiv
\omega_\epsilon^{ab}\frac{\partial(d_{\scriptscriptstyle\zeta}H_\epsilon)}
{\partial(d_{\scriptscriptstyle\zeta}z^b)}
\mathrel{\mathop=^{\rm def}}(z^a,d_{\scriptscriptstyle\zeta}H_\epsilon)_
{\epsilon+\zeta},\label{2.6}
\end{eqnarray}
where $d_{\scriptscriptstyle\zeta}$ ($\zeta=0,1$) is one of the exterior 
differentials $d_{\scriptscriptstyle0}$ or $d_{\scriptscriptstyle1}$, which 
have opposite Grassmann parities $0$ 
and $1$ respectively and following symmetry properties with respect to the 
ordinary multiplication
\begin{eqnarray}
d_{\scriptscriptstyle 0}z^ad_{\scriptscriptstyle 0}z^b=
(-1)^{g_ag_b}d_{\scriptscriptstyle0}z^bd_{\scriptscriptstyle0}z^a,\nonumber
\end{eqnarray}
\begin{eqnarray}\label{2.8}
d_{\scriptscriptstyle1}z^ad_{\scriptscriptstyle1}z^b=
(-1)^{(g_a+1)(g_b+1)}d_{\scriptscriptstyle1}z^bd_{\scriptscriptstyle1}z^a
\end{eqnarray}
and exterior products
\begin{eqnarray}
d_{\scriptscriptstyle 0}z^a\wedge d_{\scriptscriptstyle 0}z^b=
(-1)^{g_ag_b+1}d_{\scriptscriptstyle0}z^b\wedge d_{\scriptscriptstyle0}z^a,
\nonumber
\end{eqnarray}
\begin{eqnarray}\label{2.10}
d_{\scriptscriptstyle1}z^a\tilde\wedge d_{\scriptscriptstyle1}z^b=
(-1)^{(g_a+1)(g_b+1)}d_{\scriptscriptstyle1}z^b\tilde\wedge 
d_{\scriptscriptstyle1}z^a.
\end{eqnarray}
We use different notations $\wedge$ and $\tilde\wedge$ for the exterior 
products of $d_{\scriptscriptstyle 0}z^a$ and $d_{\scriptscriptstyle1}z^a$ 
respectively. 

By taking the exterior differential
$d_{\scriptscriptstyle\zeta}$ from the Hamilton equations (\ref{2.5}), we 
obtain
\begin{eqnarray}
\frac{d(d_{\scriptscriptstyle\zeta}z^a)}{dt}=(d_{\scriptscriptstyle\zeta}
\omega_\epsilon^{ab})
\frac{\partial(d_{\scriptscriptstyle\zeta}H_\epsilon)}{\partial
(d_{\scriptscriptstyle\zeta}z^b)}
+(-1)^{\zeta(g_a+\epsilon)}\omega_\epsilon^{ab}\partial_{z^b}
(d_{\scriptscriptstyle\zeta}H_\epsilon)
\mathrel{\mathop=^{\rm def}}(d_{\scriptscriptstyle\zeta}z^a,
d_{\scriptscriptstyle\zeta}H_\epsilon)_{\epsilon+\zeta}.\label{2.11}
\end{eqnarray}
As a result of equations (\ref{2.6}) and (\ref{2.11}) we have by definition 
the following binary composition for functions $F$ and $H$ of the variables 
$z^a$ and their differentials $d_{\scriptscriptstyle\zeta}z^a\equiv 
y_{\scriptscriptstyle\zeta}^a$
\begin{eqnarray}\label{2.12}
(F,H)_{\epsilon+\zeta}=
F\Bigl[\rpar_{z^a}\omega_\epsilon^{ab}\lpar_{y_{\scriptscriptstyle\zeta}^b}
&+(-1)^{\zeta(g_a+\epsilon)}\rpar_{y_{\scriptscriptstyle\zeta}^a}
\omega_\epsilon^{ab}\lpar_{z^b}\cr\nonumber\\
&+\rpar_{y_{\scriptscriptstyle\zeta}^a}y_{\scriptscriptstyle\zeta}^c
\left(\partial_{z^c}
\omega_\epsilon^{ab}\right)\lpar_{y_{\scriptscriptstyle\zeta}^b}\Bigr]H.
\end{eqnarray}
By using relations (\ref{2.2})-(\ref{2.4}) for the matrix 
$\omega_\epsilon^{ab}$, we can 
establish 
that the composition (\ref{2.12}) satisfies all the main properties
for the Poisson bracket with the Grassmann parity equal to $\epsilon+\zeta$. 
Thus, the application of the exterior differentials of opposite Grassmann
parities to the given Poisson bracket results in the brackets of the
different Grassmann parities.

By transition to the co-differential variables 
$y^{\scriptscriptstyle\epsilon+\scriptscriptstyle\zeta}_a$, related
with differentials $y_{\scriptscriptstyle\zeta}^a$ by means of the matrix 
$\omega_\epsilon^{ab}$
\begin{eqnarray}\label{2.13}
y_{\scriptscriptstyle\zeta}^a=y^{\scriptscriptstyle\epsilon+
\scriptscriptstyle\zeta}_b\omega_\epsilon^{ba},
\end{eqnarray}
the Poisson bracket (\ref{2.12}) takes a canonical form\footnote{There is 
no summation over $\epsilon$ in relation (\ref{2.13}).}
\begin{eqnarray}\label{2.14}
(F,H)_{\epsilon+\zeta}=F\left[\rpar_{z^a}
\lpar_{y^{\scriptscriptstyle\epsilon+\scriptscriptstyle\zeta}_a}
-(-1)^{g_a(g_a+\epsilon+\zeta)}\rpar_{y^{\scriptscriptstyle\epsilon+
\scriptscriptstyle\zeta}_a}\lpar_{z^a}\right]H,
\end{eqnarray}
that can be proved with the use of the Jacobi identity (\ref{2.4}).

The bracket (\ref{2.12}) is given on the functions of the variables
$z^a$, $y_{\scriptscriptstyle\zeta}^a$
\begin{eqnarray}
F=\sum_p\frac{1}{p!} y_{\scriptscriptstyle\zeta}^{a_p}\cdots 
y_{\scriptscriptstyle\zeta}^{a_1}
f_{a_1\ldots a_p}(z), \qquad 
g(f_{a_1\ldots a_p})=g_f+g_{a_1}+\cdots+g_{a_p},\nonumber
\end{eqnarray}
whereas this bracket, rewritten in the form (\ref{2.14}), is given on the 
functions of variables $z^a$ and $y^{\scriptscriptstyle\epsilon+
\scriptscriptstyle\zeta}_a$
\begin{eqnarray}
F=\sum_p\frac{1}{p!} y^{\scriptscriptstyle\epsilon+\scriptscriptstyle\zeta}_{a_p}\cdots y^{\scriptscriptstyle\epsilon+\scriptscriptstyle\zeta}_{a_1}
f^{a_1\ldots a_p}(z), \qquad 
g(f^{a_1\ldots a_p})=g_f+\epsilon p+g_{a_1}+\cdots+g_{a_p}.\nonumber
\end{eqnarray}
We do not exclude a possibility of the own Grassmann parity $g_f\equiv g(f)$ 
for a quantity $f$. 


{\bf III.\/} If we take the bracket in the canonical form (\ref{2.14}), then we obtain 
the generalizations of the Schouten-Nijenhuis bracket~\cite{scho,nij} 
(see also~\cite{nij1,fr-nij,kod-sp,but,bffls,oz,karmas}) onto the cases of 
superspace and the brackets of diverse Grassmann parities. Indeed, let us 
consider the bracket (\ref{2.14}) between monomials $F$ and $H$ having 
respectively degrees $p$ and $q$
\begin{eqnarray}
F=\frac{1}{p!} y^{\scriptscriptstyle\epsilon+\scriptscriptstyle\zeta}_{a_p}
\cdots y^{\scriptscriptstyle\epsilon+\scriptscriptstyle\zeta}_{a_1}
f^{a_1\ldots a_p}(z), \qquad 
g(f^{a_1\ldots a_p})=g_f+ p\epsilon+g_{a_1}+\cdots+g_{a_p},\nonumber
\end{eqnarray}
\begin{eqnarray}
H=\frac{1}{q!} y^{\scriptscriptstyle\epsilon+\scriptscriptstyle\zeta}_{a_q}
\cdots y^{\scriptscriptstyle\epsilon+\scriptscriptstyle\zeta}_{a_1}
h^{a_1\ldots a_q}(z), \qquad 
g(h^{a_1\ldots a_q})=g_h+q\epsilon+g_{a_1}+\cdots+g_{a_q}.\nonumber
\end{eqnarray}
Then as a result we obtain
\begin{eqnarray}
(F,H)_{\epsilon+\zeta}
&=&\frac{(-1)^{[g_{b_1}+\cdots+g_{b_{q-1}}+(q-1)(\epsilon+\zeta)]
(g_f+g_l+p\zeta)}}{p!(q-1)!}\cr\nonumber\\
&\times& y^{\scriptscriptstyle\epsilon+\scriptscriptstyle\zeta}_{b_{q-1}}
\cdots y^{\scriptscriptstyle\epsilon+\scriptscriptstyle\zeta}_{b_1}
y^{\scriptscriptstyle\epsilon+\scriptscriptstyle\zeta}_{a_p}
\cdots y^{\scriptscriptstyle\epsilon+\scriptscriptstyle\zeta}_{a_1}
\left(f^{a_1\ldots a_p}\rpar_{z^l}\right)h^{b_1\ldots b_{q-1}l}\cr\nonumber\\
&-&\frac{(-1)^{(g_l+\epsilon+\zeta)(g_f+p\epsilon+g_{a_2}+\cdots+g_{a_p})+
[g_{b_1}+\cdots+g_{b_q}+q(\epsilon+\zeta)][g_f+\epsilon+(p-1)\zeta]}}
{(p-1)!q!}\cr\nonumber\\
&\times& y^{\scriptscriptstyle\epsilon+\scriptscriptstyle\zeta}_{b_q}
\cdots y^{\scriptscriptstyle\epsilon+\scriptscriptstyle\zeta}_{b_1}
y^{\scriptscriptstyle\epsilon+\scriptscriptstyle\zeta}_{a_p}
\cdots y^{\scriptscriptstyle\epsilon+\scriptscriptstyle\zeta}_{a_2}
f^{la_2\ldots a_p}\partial_{z^l}h^{b_1\ldots b_q}.\label{2.18}
\end{eqnarray}


Let us consider the formula (\ref{2.18}) for the particular values of 
$\epsilon$ and $\zeta$. 

1. We start from the case which leads to the usual Schouten-Nijenhuis bracket 
for the skew-symmetric contravariant tensors. In this case, when $\epsilon=0$, 
$\zeta=1$ and the matrix $\omega_0^{ab}(x)=-\omega_0^{ba}(x)$ corresponds to 
the usual Poisson bracket for the commuting coordinates $z^a=x^a$, we have
\begin{eqnarray}
(F,H)_1&=&\frac{(-1)^{(q-1)(g_f+p)}}{p!(q-1)!}
\Theta_{b_{q-1}}\cdots\Theta_{b_1}\Theta_{a_p}\cdots\Theta_{a_1}
\left(f^{a_1\ldots a_p}\rpar_{x^l}\right)h^{b_1\ldots b_{q-1}l}\cr\nonumber\\
&-&\frac{(-1)^{g_f(q+1)+q(p-1)}}{(p-1)!q!}
\Theta_{b_q}\cdots\Theta_{b_1}\Theta_{a_p}\cdots\Theta_{a_2}
f^{la_2\ldots a_p}\partial_{x^l}h^{b_1\ldots b_q},\label{3.1}
\end{eqnarray}
where $\Theta_a\equiv y_a^{\scriptscriptstyle1}$ are Grassmann co-differential 
variables related owing to (\ref{2.13}) with the Grassmann differential 
variables $\Theta^a\equiv d_{\scriptscriptstyle1}x^a$
\begin{eqnarray}
\Theta^a=\Theta_b\omega_0^{ba}.\nonumber
\end{eqnarray}
When Grassmann parities of the quantities $f$ and $h$ are equal to zero
$g_f=g_h=0$, we obtain from (\ref{3.1})
\begin{eqnarray}
(F,H)_1\mathrel{\mathop=^{\rm def}}
(-1)^{(p+1)q+1}\Theta_{a_{p+q}}\cdots\Theta_{a_2}[F,H]^{a_2\ldots a_{p+q}},
\nonumber
\end{eqnarray}
where $[F,H]^{a_2\ldots a_{p+q}}$ are components of the usual 
Schouten-Nijenhuis bracket (see, for example, \cite{bffls}) for the 
contravariant antisymmetric tensors\footnote{Here and below we use the same 
notation [F,H] for the different brackets. We hope that this will not lead to 
the confusion.}. 
This bracket has the following properties
\begin{eqnarray}
[F,H]=(-1)^{pq}[H,F],\qquad
\sum_{(FHE)}(-1)^{ps}[[F,H],E]=0,\nonumber
\end{eqnarray}
where $s$ is a degree of a monomial $E$.

2. In the case $\epsilon=\zeta=0$ and $\omega_0^{ab}(x)=-\omega_0^{ba}(x)$ we 
obtain the 
bracket for symmetric contravariant tensors (see, for example, \cite{but})
\begin{eqnarray}
(F,H)_0
\mathrel{\mathop=^{\rm def}}
y^{\scriptscriptstyle0}_{a_{p+q}}\cdots y^{\scriptscriptstyle0}_{a_2}
[F,H]^{a_2\ldots a_{p+q}},\nonumber
\end{eqnarray}
where commuting co-differentials $y^{\scriptscriptstyle0}_a$ connected with 
commuting differentials $y^a_{\scriptscriptstyle0}\equiv 
d_{\scriptscriptstyle0}x^a$ in accordance with (\ref{2.13})
\begin{eqnarray}
y_{\scriptscriptstyle0}^a=y^{\scriptscriptstyle0}_b\omega_0^{ba}\nonumber
\end{eqnarray}
and the bracket $[F,H]^{a_2\ldots a_{p+q}}$ has the following  
properties
\begin{eqnarray}
[F,H]=-(-1)^{g_fg_h}[H,F],\qquad
\sum_{(EFH)}(-1)^{g_eg_h}[E,[F,H]]=0.\nonumber
\end{eqnarray}

3. By taking the Martin bracket \cite{mar} $\omega_0^{ab}(\theta)=
\omega_0^{ba}(\theta)$ with Grassmann coordinates $z^a=\theta^a$ $(g_a=1)$ as 
an initial bracket (\ref{2.1}), we have in the case $\zeta=0$ for 
antisymmetric contravariant tensors on the Grassmann algebra
\begin{eqnarray}
(F,H)_0\stackrel{\rm def}{=}
\Theta_{a_{p+q}}\cdots\Theta_{a_2}[F,H]^{a_2\ldots a_{p+q}},\nonumber
\end{eqnarray}
where the Grassmann co-differentials $\Theta_a$ related with the Grassmann 
differentials $\Theta^a$ as
\begin{eqnarray}
d_{\scriptscriptstyle0}\theta^a\equiv\Theta^a=\Theta_b\omega_0^{ba}.\nonumber
\end{eqnarray}
The bracket $[F,H]$ has the following properties
\begin{eqnarray}
[F,H]=-(-1)^{g_fg_h}[H,F],\qquad
\sum_{(EFH)}(-1)^{g_eg_h}[E,[F,H]]=0.\nonumber
\end{eqnarray}

4. By taking the Martin bracket again, in the case $\zeta=1$
\begin{eqnarray}
d_{\scriptscriptstyle1}\theta^a\equiv x^a=x_b\omega_0^{ba}\nonumber
\end{eqnarray}
we obtain for the symmetric tensors on Grassmann algebra
\begin{eqnarray}
(F,H)_1\stackrel{\rm def}{=}
x_{a_{p+q}}\cdots x_{a_2}[F,H]^{a_2\ldots a_{p+q}}.\nonumber
\end{eqnarray}
The bracket $[F,H]$ has the following properties
\begin{eqnarray}
[F,H]=-(-1)^{(g_f+p+1)(g_h+q+1)}[H,F],\qquad
\sum_{(EFH)}(-1)^{(g_e+s+1)(g_h+q+1)}[E,[F,H]]=0.\nonumber
\end{eqnarray}

5. In general, if we take the even bracket in superspace with coordinates 
$z^a=(x,\theta)$, where $x$ and $\theta$ are respectively commuting and 
anticommuting (Grassmann) variables, then in the case $\zeta=1$ we have
\begin{eqnarray}
(F,H)_1\stackrel{\rm def}{=}
y^{\scriptscriptstyle1}_{a_{p+q}}\cdots y^{\scriptscriptstyle1}_{a_2}
[F,H]^{a_2\ldots a_{p+q}},\nonumber
\end{eqnarray}
where
\begin{eqnarray}
d_{\scriptscriptstyle1}z^a\equiv y^a_{\scriptscriptstyle1}=
y^{\scriptscriptstyle1}_b\omega_0^{ba}.\nonumber
\end{eqnarray}
The bracket $[F,H]$ has the following properties
\begin{eqnarray}
[F,H]=-(-1)^{(g_f+p+1)(g_h+q+1)}[H,F],\qquad
\sum_{(EFH)}(-1)^{(g_e+s+1)(g_h+q+1)}[E,[F,H]]=0.\nonumber
\end{eqnarray}

6. In the case of the even bracket in superspace as initial one with $\zeta=0$ 
we obtain
\begin{eqnarray}
(F,H)_0\stackrel{\rm def}{=}
y^{\scriptscriptstyle0}_{a_{p+q}}\cdots y^{\scriptscriptstyle0}_{a_2}
[F,H]^{a_2\ldots a_{p+q}},\nonumber
\end{eqnarray}
where
\begin{eqnarray}
d_{\scriptscriptstyle0}z^a\equiv y^a_{\scriptscriptstyle0}=
y^{\scriptscriptstyle0}_b\omega_0^{ba}.\nonumber
\end{eqnarray}
The bracket $[F,H]$ has the following properties
\begin{eqnarray}
[F,H]=-(-1)^{g_fg_h}[H,F],\qquad
\sum_{(EFH)}(-1)^{g_eg_h}[E,[F,H]]=0.\nonumber
\end{eqnarray}

7. Taking as an initial bracket the odd Poisson bracket in superspace with 
coordinates $z^a$, for the case $\zeta=0$ we have
\begin{eqnarray}
(F,H)_1
\stackrel{\rm def}{=}
y^{\scriptscriptstyle1}_{a_{p+q}}\cdots y^{\scriptscriptstyle1}_{a_2}
[F,H]^{a_2\ldots a_{p+q}},\nonumber
\end{eqnarray}
where
\begin{eqnarray}
d_{\scriptscriptstyle0}z^a\equiv y^a_{\scriptscriptstyle0}=
y^{\scriptscriptstyle1}_b\omega_1^{ba}.\nonumber
\end{eqnarray}
The bracket $[F,H]$ has the following properties
\begin{eqnarray}
[F,H]=-(-1)^{(g_f+1)(g_h+1)}[H,F],\qquad
\sum_{(EFH)}(-1)^{(g_e+1)(g_h+1)}[E,[F,H]]=0.\nonumber
\end{eqnarray}

8. At last for the odd Poisson bracket in superspace, taking as an initial 
one, we obtain in the case $\zeta=1$
\begin{eqnarray}
(F,H)_0\stackrel{\rm def}{=}
y^{\scriptscriptstyle0}_{a_{p+q}}\cdots y^{\scriptscriptstyle0}_{a_2}
[F,H]^{a_2\ldots a_{p+q}},\nonumber
\end{eqnarray}
where
\begin{eqnarray}
d_{\scriptscriptstyle1}z^a\equiv y^a_{\scriptscriptstyle1}=
y^{\scriptscriptstyle0}_b\omega_1^{ba}.\nonumber
\end{eqnarray}
The bracket $[F,H]$ has the following  
properties
\begin{eqnarray}
[F,H]=-(-1)^{(g_f+p)(g_h+q)}[H,F],
\qquad \sum_{(EFH)}(-1)^{(g_e+s)(g_h+q)}[E,[F,H]]=0.\nonumber
\end{eqnarray}

Thus, we see that the formula (\ref{2.18}) contains as particular cases quite 
a number of the Schouten-Nijenhuis type brackets.


{\bf IV.\/} We give the prescription for 
the construction from a given Poisson bracket of the definite Grassmann parity 
another bracket. For this construction we use the exterior differentials with 
different Grassmann parities. We proved  that the resulting Poisson bracket 
essentially depends on the parity of the exterior differential in spite of 
these differentials give the same exterior calculus 
\cite{Soroka:soroka&soroka}. The 
prescription leads to the set of different generalizations for the 
Schouten-Nijenhuis bracket. Thus, we see that the Schouten-Nijenhuis bracket 
and its possible generalizations are particular cases of the usual Poisson 
brackets of different Grassmann parities (\ref{2.14}). We hope that these 
generalizations will find their own application for the deformation 
quantization (see, for example, \cite{bffls,konts}) as well as the usual 
Schouten-Nijenhuis bracket.

We are sincerely grateful to J. Stasheff for the interest to the work and 
stimulating remarks. One of the authors (V.A.S.) sincerely thanks L. Bonora 
for the fruitful discussions and warm hospitality at the SISSA/ISAS (Trieste) 
where this work has been completed.

\end{document}